\RequirePackage[2020-02-02]{latexrelease}
\documentclass[aps,prd,floatfix,superscriptaddress,onecolumn,amssymb]{revtex4}

\usepackage{amssymb,amsmath,verbatim}
\usepackage{mathtools}
\usepackage{graphicx}
\usepackage{epsfig}
\usepackage{dcolumn}
\usepackage{bm}
\usepackage{epstopdf}
\usepackage{float}
\usepackage{xcolor}

\newcommand{\be}{\begin{equation}}
\newcommand{\ee}{\end{equation}\noindent}
\newcommand{\eei}{\end{equation}}
\newcommand{\bea}{\begin{eqnarray}}
\newcommand{\eea}{\end{eqnarray}\noindent}
\newcommand{\eeai}{\end{eqnarray}}

\newcommand{\hf} {\frac{1}{2}}
\newcommand{\nn}{\nonumber\\}
\newcommand\eqn[1]{Eq.\,(\ref{#1})}

\newcommand\fig[1]{Fig.\,{\ref{#1}}}
\newcommand\sect[1]{Sect.\,{\ref{#1}}}

\def\eq#1{(\ref{#1})}

\def\ord#1{{\cal O}(#1)}

\def\h#1{{\hat#1}}

\def\c#1{{\cal#1}}
\def\b#1{{\bar#1}}

\def\Tr{{\mathrm{Tr}}}
\def\v#1{{\bm{#1}}}

\begin{document}
\title{On the Lorentz symmetry in conformally reduced Quantum Gravity}

\author{F. Gégény}
\affiliation{Department of Theoretical Physics, University of Debrecen,
	P.O. Box 5, H-4010 Debrecen, Hungary}
\author{S. Nagy}
\affiliation{Department of Theoretical Physics, University of Debrecen,
P.O. Box 5, H-4010 Debrecen, Hungary}
\author{K. Sailer}
\affiliation{Department of Theoretical Physics, University of Debrecen,
P.O. Box 5, H-4010 Debrecen, Hungary}

\date{\today}

\begin{abstract}

The functional renormalization group treatment of the conform reduced Einstein-Hilbert gravity is extended by following the evolution of the time and space derivatives separately, in order to consider the Lorentz symmetry during the evolution. We found the Reuter fixed point in the ultraviolet region. It is shown that starting from the Gaussian fixed point the Lorentz symmetry breaks down in the vicinity of the Reuter fixed point. Similarly, in the symmetry broken phase it also breaks down in the infrared region close to a critical  singularity scale. By calculating the anomalous dimension form the kinetic term of the action, we found a new relevant coupling belonging to the curvature.

\end{abstract}


\maketitle

\section{Introduction}

The unification of quantum mechanics and relativity is one of the greatest challenges in modern physics. Although there are several promising attempts to combine the two theories consistently, the full success is waiting to happen. Among the many trials we should mention loop quantum gravity \cite{Rovelli:1989za,Rovelli:1997yv,Thiemann:2002nj,Ashtekar:2004eh}, string theory \cite{Green:1987sp,Polchinski:1998rq}, and the asymptotically safe (AS) gravity \cite{Reuter:2007rv,Dupuis:2020fhh,Percacci:2017fkn,Reuter:2019byg}, where the Newton- and the cosmological constants run, and the metric plays the role of the field variables and stands for the gravitational interactions. The quantum mechanical description of elementary particles leads us to the Standard Model (SM), which is one of the greatest achievements of modern physics. The Lorentz group and the Lorentz-symmetry play crucial roles in building up the SM, therefore the investigation of Lorentz invariance is fundamental in theories, where the unification has been made.

It was a great success of the functional renormalization group (RG) method \cite{Wetterich:1992yh,Morris:1994ki,Polonyi:2001se,Pawlowski:2005xe,Dupuis:2020fhh} to find an interacting, non-Gaussian fixed point (NGFP), called Reuter fixed point \cite{Reuter:1996cp} in gravity, which makes the model asymptotically safe. During the RG procedure we remove modes from the system, and push them to the environment in a systematic way, and it can follow the energy dependence of the models. Usually, the method starts from an ultraviolet cutoff and follows the theory to lower energies. However, in AS gravity we should perform the RG blocking up into the ultraviolet (UV) direction, since the Reuter fixed point is UV attractive. Recently, one of the most important issues for the RG method was to show that the gravity models and their extensions or generalizations remain asymptotically safe \cite{Dona:2013qba,Eichhorn:2017lry,Moti:2018rho,Gubitosi:2019ncx,DeBrito:2019rrh,Laporte:2021kyp,Daas:2021abx,Ferrero:2022hor,Ferrero:2022dpk}.

Usually the RG method is formulated in Euclidean spacetime, however the proper treatment requires a formulation with Lorentz signature. The formulation of gravitational interaction and the corresponding RG equations using Lorentzian signature has been widely investigated in the literature. Various approaches have been considered to formulate the RG method with Lorentz signature, e.g. the method of causal sets \cite{Eichhorn:2017bwe, Eichhorn:2019xav}, the Arnowitt-Deser-Misner (ADM) decomposition \cite{Manrique:2011jc,Rechenberger:2012dt,Biemans:2016rvp,Biemans:2017zca,Houthoff:2017oam,Platania:2017djo,Baldazzi:2019kim}, where the time direction is singled out.

The Lorentz symmetry is thought to be an emergent symmetry at low energies in the IR, however it is broken in the UV around the Planck scale. The Lorentz-symmetry, and its possible violation, has been extensively explored \cite{Jacobson:2000xp,Carroll:2001ws,Horava:2009uw,Liberati:2013xla,Eichhorn:2019ybe}, in the matter and in the pure gravitational sectors. The violation is investigated from a different aspect in \cite{Knorr:2018fdu} where the diffeomorphism symmetry of the RG equations for foliated spacetimes are kept, but Lorentz-symmetry breaking terms are introduced into the action, and according to the proposal, the Lorentz symmetry is violated in the vicinity of the UV Reuter fixed point. The Lorentzian treatment plays an important role in several other works, e.g. in \cite{Fehre:2021eob} where the flow equation for the graviton spectral function in Lorentzian signature is postulated, in \cite{Bonanno:2021squ} where the graviton propagator in AS quantum gravity is reconstructed from Euclidean data, in \cite{Knorr:2021niv} where the momentum-dependent form factors are determined in $\varGamma$ from first principles, or \cite{Houthoff:2020zqy} where composite operators of the AS gravity are considered.

The shortcoming of the renormalization is that the Lorentz signature can only be adopted with great difficulties. On the one hand, in quantum field theories we need regularizations, e.g. in the Wetterich equation, which controls the evolution of the effective action, the regulator plays a prominent role. However, up to our present knowledge, the nonperturbative regulators are not Lorentz invariant \cite{Polonyi:2018ykh}. On the other hand, the renormalization procedure usually breaks the Lorentz symmetry itself, for example a possible way of performing the renormalization is to take the gliding cutoff scale $k$ from the length of the spatial momentum space, while the temporal momentum is integrated out \cite{Steib:2019xrv}. We follow the same procedure in this work, too. Using such an RG scheme is a rather compromised soulution, since it breaks the Lorentz symmetry, but it can be used very effectively. To enlighten the problem of the RG blocking we note that the loop integral in the Wetterich equation is a momentum integral on a $d$-dimensional sphere with finite volume in Euclidean formalism, however it is an infinite volume hyperboloid in Minkowski spacetime, and the latter needs a further regularization. 

The Euclidean and the Lorentz formalisms differ significantly. The Euclidean propagator can describe off-shell, virtual particles, while the propagator in Minkowski spacetime can have on-shell contributions from the real particles, which are missing in the former case. Furthermore the Minkowski formalism makes the couplings complex, that can result in complex fixed points and exponents. The standard Wick rotation cannot guarantee the equivalence of the two formulations, since it breaks the Lorentz symmetry, too, furthermore the dynamical nature of the metric also influences this issue \cite{Bonanno:2021squ,Fehre:2021eob}.

These reasons lead us to choose such an RG study, where we integrate out the temporal momentum separately and consider only the spatial momenta during the RG evolution, which is then finite. Fortunately, the results show that such regulators and cutoffs, that break the Lorentz symmetry, lead to such evolution into the IR, that keeps the Lorentz symmetry, showing that our treatment is reliable.

We apply the RG method to the conformally reduced model of quantum gravity. We introduce a new curvature coupling in the Einstein-Hilbert action, which proved to be relevant. Furthermore we take the temporal and spatial momenta in the kinetic term in an anistropic way. Temporal momentum is multiplied by a further coupling, and its evolution can signal the possible Lorentz-symmetry violation. We found that this coupling is also relevant. The phase space completed by these new couplings also possesses the UV Reuter fixed point, moreover, it also shows another, "high UV" fixed point that can be viewed as the migration of the original Reuter fixed point due to the new coupling. Furthermore, the Lorentz symmetry is violated in the UV region starting in the vicinity of the Reuter fixed point in accordance with the aforementioned arguments.

The paper is organized as follows. In \sect{sect:asg} we treat the evolution equation of AS gravity in Minkowski spacetime and derive the beta functions. In \sect{sect:rel} we look for the relevant couplings and the structure of the phase space. We also investigate the UV and the infrared (IR) limits of the model. Finally, in \sect{sect:sum} the conclusions are drawn up.

\section{Asymptotically safe gravity}\label{sect:asg}

The form of the conformally reduced Einstein-Hilbert (CREH) action is
\be
S_k[f; \chi_B] = - Z \int_x \sqrt{ \h{g}_x} \left\{ - \hf f  \h{\Box} f  + \frac{1}{12} \h{R} (\chi_B +f)^2 - \frac{\Lambda}{6}  (\chi_B +f)^4 \right\}.
\ee
In the derivation of this CREH action it has been assumed that the background field $\chi_B=$const. The (blocked) action depends on the renormalization scale $k$. Newton's gravitational coupling $G=G_k$ appears in the wavefunction renormalization $Z= \frac{3}{4\pi G}$, and $\Lambda=\Lambda_k$ stands for the  cosmological "constant''. For simpler notations, we suppress the scale-dependence of the couplings. Here $\h{g}$ is the reference metric, and $-\h{\Box}$ is the Beltrami-Laplace operator. A similar form of the effective action can be assumed
\be\label{ansatz}
 \Gamma_k[\b{f}; \chi_B] = - Z_k \int_x \sqrt{ \h{g}_x} \left\{ - \hf \b{f}  \h{\Box} \b{f}  + \frac{c}{12} \h{R} (\chi_B +\b{f})^2 - \frac{\Lambda}{6}  (\chi_B +\b{f})^4 \right\} 
\ee
 in the lowest order of the derivative expansion. The coupling $c$ has been introduced additionally, that enables one to run the derivative term and the $\ord{\h{R}}$ term separately. If we set $c=1$ and it does not run, we can reproduce the usual 2-coupling conformally reduced version of the Einstein-Hilbert action. For cylindrical geometry we have
\bea
-\h{\Box}&=& W\partial_0^2 -\h{\Delta} 
\eea
when the Lorentz signature is used. We multiply the time derivative with an additional running coupling $W$. When $W\ne 1$ then the Lorentz symmetry is violated. The background metric to the running scale $k$ is $\b{g}_{\mu\nu} = \chi_B^2\h{g}_{\mu\nu}$. The regulator should be introduced in order to regulate the momentum integrals around the scale $k^2$ of the eigenvalue of the operator $-\b{\Box}$ corresponding to the background metric $ \b{g}_{\mu\nu}$ and implying the replacement
\be\label{regul}
 -\b{\Box} \Rightarrow  -\b{\Box} + k^2R^{(0)} \biggl(\frac{-\h{\Box}}{k^2\chi_B^2} \biggr),
\ee
where $\h\Box=\chi_B^2\b\Box$ is used \cite{Wetterich:1992yh,Niedermaier:2006wt}. The choice of the  reference metric depends on the signature and on the manner how the frequency and 3-momentum integrals are performed when the trace on the r.h.s. of the Wetterich equation is evaluated.  For the  Lorentz signature we choose the reference metric $\h{g}_{00}=-1$ and $\h{g}_{ij}=\delta_{ij}$ $(i,j=1,2,3)$. The temporal momentum integral is performed separately and the spatial momentum integral is performed by means of the heat-kernel approach applied to functions of the 3-dimensional $-\h{\Delta}$ operator. The regulated operator becomes
\be
-\b{\Box} + \gamma  k^2R^{(0)}  \biggl(\frac{-\h{\Delta}}{k^2\chi_B^2} \biggr) = \chi_B^{-2} \biggl\lbrack
 W \partial_0^2 -\h{\Delta} +    \gamma \chi_B^2 k^2R^{(0)}  \biggl(\frac{-\h{\Delta}}{k^2\chi_B^2} \biggr)\biggr\rbrack.
 \ee
In the path integral the regulation the replacement \eq{regul} corresponds to the change
\be
- Z\int_x \hf f (-\h{\Box}) f \Rightarrow - Z\int_x \hf f\biggl\lbrack W\partial_0^2 -\h{\Delta} +    (1-i\epsilon) \chi_B^2 k^2R^{(0)}  \biggl(\frac{-\h{\Delta}}{k^2\chi_B^2} \biggr)\biggr\rbrack f.
\ee
We introduce the notation for the regulator
\be
\c{R}_k=- Z (1-i\epsilon) \chi_B^2 k^2R^{(0)}  \biggl(\frac{-\h{\Delta}}{k^2\chi_B^2} \biggr).
\ee
The $-i\epsilon$ term makes the path integral convergent and the momentum-dependence of the inverse propagator is cancelled for the Litim regulator. This implies that the inverse propagator in momentum space is
\be
-W\omega^2 +\v{p}^2 + (1-i\epsilon)\chi_B^2 k^2R^{(0)}  \biggl(\frac{\v{p}^2}{\chi_B^2 k^2} \biggr) =- \chi_B^2 k^2 \biggl\lbrack W s^2-\biggl(1 -i\epsilon(1-z) \biggr)\biggr\rbrack,
\ee
where we used the form of the Litim regulator and the dimensionless variables $s^2=\frac{\omega^2}{\chi_B^2k^2}$ and $z=\frac{\v{p}^2}{\chi_B^2k^2}$. It provides the propagator $ \lbrack W s^2 - (1-i\epsilon) \rbrack^{-1}$, with the effective mass term $m^2_{eff}=W^{-1}(1-i\epsilon)$, which is in accordance with the Feyman's $i\epsilon$ prescription. For simplicity we suppress the $i\epsilon$ terms below.

\subsection{Evolution equations}

The Wetterich equation with Lorentz signature is
\bea\label{WEL}
 {\dot{ \Gamma}}_k[\b{f};\chi_B] &=& 
 - i\hf \Tr \biggl( \lbrack \Gamma_k^{(2)}+ \c{R}_k \rbrack^{-1}  {\dot{\c{R}}}_k\biggr),
\eea
with $\Tr A = \int_x A_{xx}=\int d^4 x \sqrt{|\h{g}_x|} A_{xx} = \int d^4x d^4 y \delta_{x,y} \sqrt{|\h{g}_x|} A_{xy} $, $\Gamma_k^{(2)}$ is the second functional derivative of the effective action with respect to the field variable. The dot stands for $\partial_t$, where $t=\ln k$.
We introduce the potentials
\be
V(\phi)= Z U(\phi),~~U(\phi)= \frac{\Lambda}{6} \phi^4-\frac{c}{12}\h{R}\phi^2,~~u(\phi)= U(\phi)+\frac{c}{12}\h{R}\phi^2.
\ee
Keeping only the $\ord{\bar f^2}$ terms, we get
\be
\dot \Gamma = -i\frac{ k^2\chi_B^2}{Z_k} ( T_0 + T_1+T_2),
\ee
where
\bea
  T_0&=& \Tr \{  K^{-1}\c{N}\},\\
  T_1 &=& \Tr \{ K^{-1}\delta \mu K^{-1} \c{N} \},\\
 T_2 &=& \Tr  \{ K^{-1} \delta \mu K^{-1} \delta \mu K^{-1} \c{N} \},
\eea
and
\bea
  K&=& \c{A}(\partial_0^2,-\h{\Delta} ) + \frac{c_k}{6}\h{R} \\
 \c{A}(\partial_0^2,-\h{\Delta} )&=& W\partial_0^2 -\h{\Delta} +\gamma  k^2 \chi_B^2 R^{(0)}\bigl(  \frac{-\h{\Delta} }{k^2\chi_B^2} \bigr)
 - u_k''(\chi_B)\\
 \c{N}&=& 
\biggl(1-\hf \eta \biggr) R^{(0)}\bigl(  \frac{-\h{\Delta} }{k^2\chi_B^2} \bigr) 
 -  \frac{-\h{\Delta} }{k^2\chi_B^2} R^{(0)\prime}\bigl(  \frac{-\h{\Delta} }{k^2\chi_B^2} \bigr),
 \eea
where we have introduced the anomalous dimension $\eta ={\dot G}/G$. We assume that in the cylindrical geometric layout the refrence metric has the form
\bea
 ( \h{g}_{\mu\nu}) &=&\begin{pmatrix}
     1 & 0 \cr
     0 & \h{h} \end{pmatrix},
\eea
where $\h{h}_{ij}$ does not depend on the coordinate $x^0$. The structure of the reference metric implies that the spacetime is foliated by hypersurfaces $\Sigma_t$ $(t=x^0)$ with normal vectors $n^\mu=(1,0,0,0)$, such that the external curvature of $\Sigma_t$ is zero, therefore the 4-dimensional and the 3-dimensional curvatures belonging to the reference metric are equal \cite{Nagy:2019jef}.

The flow equations for the potential $u(\chi_B)$ are obtained keeping the terms up to the order $\h{R}^1$. The initial condition at the scale $k_i$ corresponds to the CREH action with the bare couplings $\Lambda({k_i})\equiv \Lambda_i$, $c({k_i})=W({k_i})=1$ and $G({k_i}) = G_i$. The flow equations for the potential are obtained by taking the terms of the order $\b{f}^0$ on both sides of \eqn{WEL}, i.e., setting $\b{f}=0$:
\be
 \left( -\eta_N U(\chi_B) + {\dot U}(\chi_B) \right)\Omega=\frac{k^2\chi_B^2}{Z} T_0
\ee
The comparison of the terms of the various powers of $\h{R}$ of both sides provides the flow equations
\bea\label{lambdaflowLlit}
{\dot \lambda} &=& (-2+\eta) \lambda -i \frac{g}{\pi^{1/2}W^{1/2}} \left( 1 -\frac{\eta}5\right) \phi^1_{3/2}(-2\lambda),\\
\label{cflowLlit}
{\dot c}&=& \eta c +i \frac{g}{3\pi^{1/2}W^{1/2}} \left[\left(1-\frac{\eta}{3}\right)  \phi^1_{1/2}(-2\lambda)-c \left( 1 -\frac{\eta}{5}\right) \phi^2_{3/2}(-2\lambda) \right],
\eea
where the regulator functions in case of Minkowski spacetime
\bea
\phi_\nu^1(w)&=& i\frac{1}{2\Gamma(\nu+1)\sqrt{1+w}},\nn
\phi_\nu^2(w)&=& i\frac{1}{4\Gamma(\nu+1)(1+w)^{3/2}}
\eea
are introduced.

To get the formula for the anomalous dimension, we should consider the $\ord{\bar f^2}$ terms in \eqn{WEL},
\be
\int\sqrt{g}\left\{\hf(\eta W-\dot W)\b{f}\partial_0^2\b{f}+\hf\eta \b{f}\Delta \b{f}\right\}=-\frac{16\Lambda^2\chi_B^4k^2}{Z_k}\Tr(\c{N}\c{A}^{-2}\b{f} \c{A}^{-1}\b{f} )
\ee
After straightforward calculations we get the following expression of $\eta$
\be
\eta =-\frac{10\lambda^2 g}{9\pi W^{1/2}(1-2\lambda)^{7/2}},
\ee
and the evolution equation of $W$ becomes
\be
\dot W = - \frac{32\lambda^2 4\pi g}{3}\left(\frac{\eta}5-1\right)\frac{5W^{1/2}}{384\pi^2(1-2\lambda)^{7/2}}+\eta W.
\ee

The first two equations decouple from the others. There is a Reuter fixed point at $g^{\prime \star}=25.4$ and $\lambda^*=0.21$, with the exponents $s_{1,2}=-5.16\pm i6.17$. The corresponding value of $c$ is $c^*=-0.25$ with exponent $s_3=-2.35$. The fixed point value for $W$ is $W^*=0$ with $s_4=-0.62$.

\section{Relevant couplings}\label{sect:rel}

The evolution equations describe the scale dependence of the effective action, including the evolution of the couplings. In the vicinity of the fixed points the scaling of the trajectories can change. Around the fixed points, the couplings have certain scaling properties, usually they vary according to $g\sim e^{st}$, where the exponent $s$ characterizes the scaling nature of the coupling, and $t=\ln k$. The negative real part of $s$ corresponds to relevant scaling, the other sign denotes irrelevance. At the Reuter fixed point all couplings are relevant, making the theory asymptotically safe.

In our investigation we introduced two new couplings, therefore first we should check what happens to the traditional fixed points. The $\beta$ functions are analytic, they are
\bea
\dot g' &=& (2+\eta)g'-\frac{5g^{\prime 2}\lambda^2(2g'\lambda^2-9(1-2\lambda)^{7/2}\pi)}{81(1-2\lambda)^7 \pi^2}\nn
\dot \lambda &=& (-2+\eta)\lambda-\frac{g'(\eta-5)}{15\pi^{5/2}(1-2\lambda)^{1/2}}\nn
\dot W &=& \eta W-\frac{g' \lambda^2 W(\eta-5)}{9\pi(1-2\lambda)^{7/2}}\nn
\dot c &=& \eta c - g'\frac{c(\eta-5)-5(\eta-3)(1-2\lambda)}{90\pi^{5/2}(1-2\lambda)^{3/2}},
\eea
after introducing $g'=g/W^{1/2}$. First we note, that the first two equations decouple from the others. The rescaled coupling $g'$ with $\lambda$ exhibits the Reuter fixed point at $g^{\prime \star}=25.4$ and $\lambda^*=0.21$, with the exponents $s_{1,2}^r=-5.16\pm i6.17$. The corresponding value of $c$ is $c^*=-0.25$ with exponent $s_3^r=-2.35$. The fixed point value for $W$ is $W^*=0$ with $s_4^r=-0.62$, here the superscript stands for the corresponding fixed point. The real parts of the exponents are negative, which means that the couplings are relevant in the UV region. The exponents are complex conjugate to each other, resulting in spiral trajectories. 
The two new couplings prove to be relevant in the UV region, therefore the theory remains renormalizable. The existence of the Reuter fixed point is one of our most important results. The existence of the non-Gaussian fixed point (GFP) is responsible for the two phases of the model.  The GFP corresponds to a free theory. Furthermore, according to the concrete value of the cosmological and Newton couplings, our present world is situated very close to the GFP, which makes the investigation of this fixed point extremely important. In the GFP we have couplings with relevant and irrelevant scaling behaviors. According to the sign of the coupling $\lambda$ we can distinguish a symmetric phase, where $\lambda<0$ in the IR limit. The  trajectories of the other, the broken symmetric phase run at positive values of $\lambda$. 
In the broken phase the trajectories run into singularity at $\lambda=0.5$. This property of the broken phase also appears in scalar models, too \cite{Nagy:2009pj,Nagy:2012ef}. Although there are renormalization schemes where the trajectories survive the singularity, but it is not typical. We also note, that by improving the potential with higher order terms, the singular behavior might also disappear, but we argue in \cite{Pangon:2009wk,Pangon:2009pj}, that the singularity is the inherent property of the models with physical content. The inclusion of the new couplings might have resulted in further phases, however we could not find any new ones. 

We numerically determined the flows of the couplings in the UV and in the IR regions. The positive values of $t$ provide the UV scalings, while the negative values of $t$ give the IR scalings. In the IR limit we should consider both the symmetric and the broken phase trajectories. In the UV limit, it is not necessary to distinguish the phases, since it is not sensitive to the phases there. Where it was possible we plotted the UV and IR scalings together. We solved the RG equations starting from close to the GFP. On the one hand, we followed the traditional RG procedure and did the blocking into the IR direction. On the other hand, we should make the RG blocking into the UV direction, too. Although it is accepted in the RG treatments, we should note, that it can be maintained if irrelevant couplings are not included, since they diverge in the UV regime. This condition is satisfied in our model, so the UV RG blocking can provide us correct results.

We start the evolution close to the GFP. In its vicinity the interactions are weak, thus the kinetic term dominates the RG evolution. As we mentioned, our present world can be found there. We can say that we have a flat Minkowski spacetime, so we chose e.g. $W=1$ initially. 

\begin{figure}[H]
	\centering
	\includegraphics[width=0.45\linewidth]{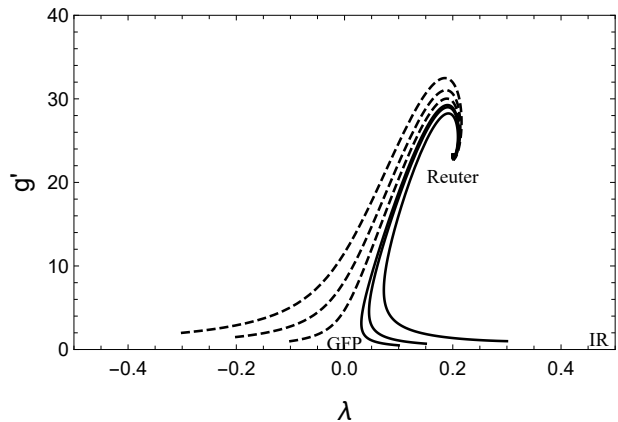}
	\caption{The $\lambda-g'$ phase diagram is presented. It shows the Gaussian, the Reuter and an infrared fixed points and two phases. The trajectories going to the left belong to the symmetric phase (dashed lines), the others are parts of the broken phase (solid lines).}
	\label{fig:lgphase}
\end{figure}

\begin{figure}[H]
	\centering
	\includegraphics[width=0.45\linewidth]{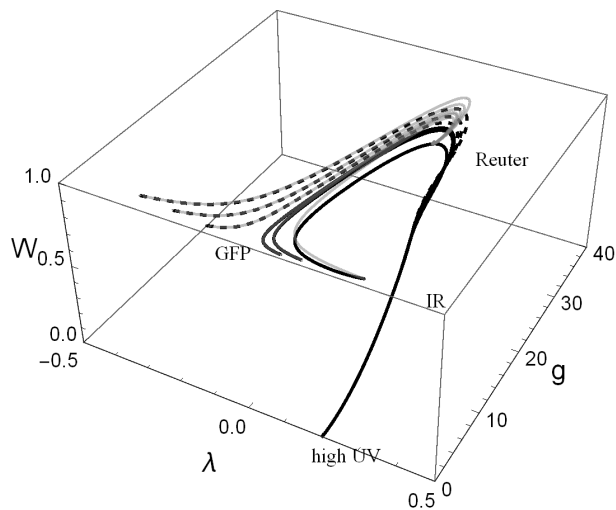}
	\caption{The $\lambda-g-W$ phase diagram is shown. It presents how the Reuter fixed point is deformed in ultraviolet region due to the coupling $W$. The trajectories of \fig{fig:lgphase} is presented on the $\lambda-g$ plane with the choice $W=1$ with grey opaque curves. The black trajectories going to the left belong to the symmetric phase (dashed lines), the others are parts of the broken phase (solid lines). In the direction of the UV they tend to another fixed point.}
	\label{fig:lgwp}
\end{figure}

\begin{figure}[H]
	\centering
	\includegraphics[width=0.45\linewidth]{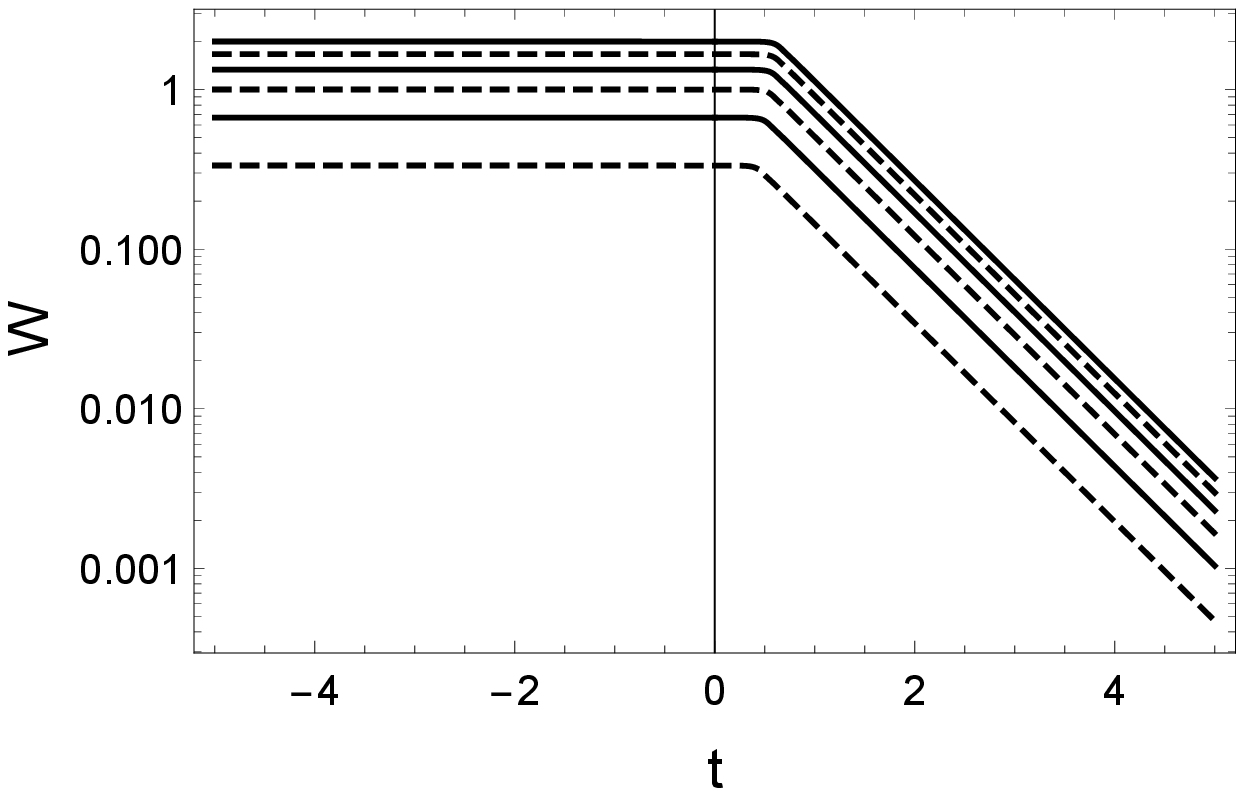} \includegraphics[width=0.45\linewidth]{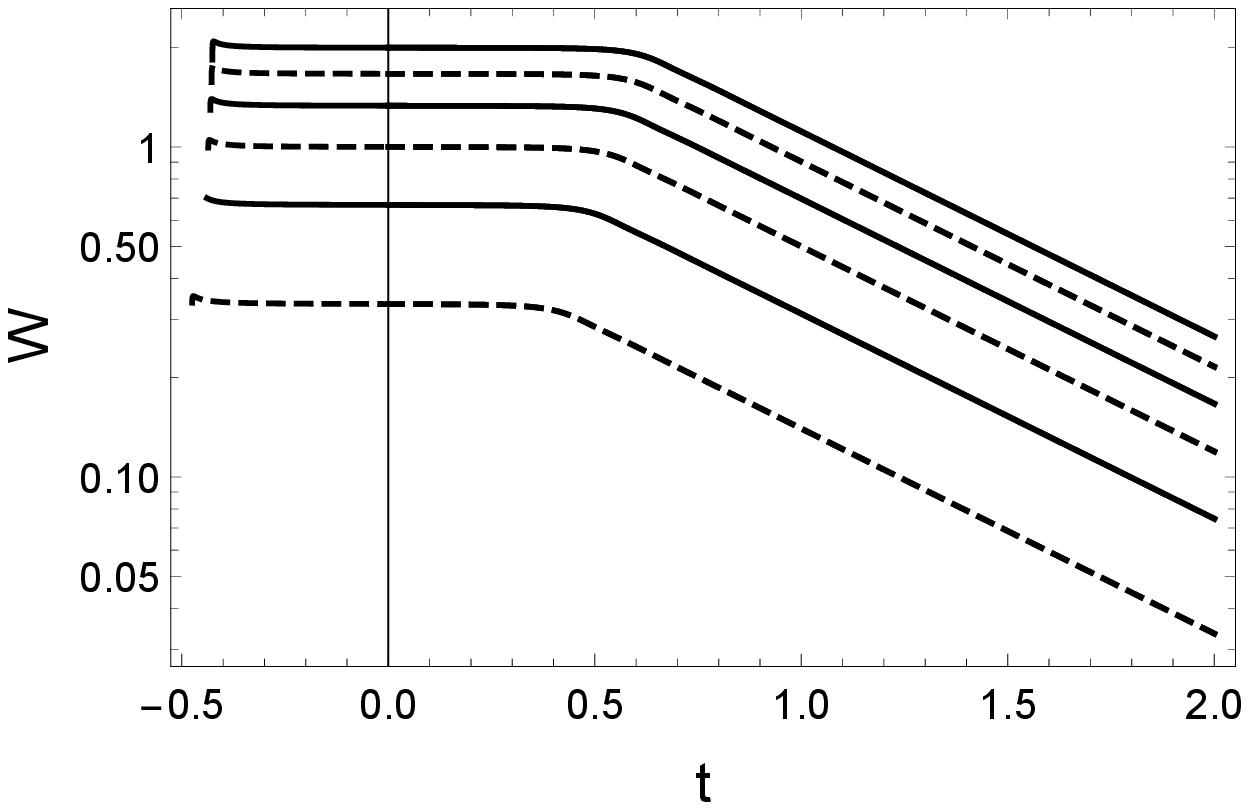}
	\caption{The left figure shows the flow of $W$ in the symmetric phase, it tends to zero in the UV limit, and it is constant from near the Gaussian fixed point towards the IR limit. The initial  values $W(k_i)=W_i$ from bottom to top are: 1/3, 2/3, 1, 4/3, 5/3, 2. The right figure shows the evolution of $ W$ in the broken symmetric phase with the same initial conditions. The UV scaling does not change, there occurs again a wide crossover region starting from near the Gaussian fixed point, where $W$ keeps  a constant value, but $W$  drops to zero suddenly  close to the singularity.}
	\label{fig:wsym}
\end{figure}

The phase space exhibiting the Reuter fixed point with spiral trajectories is shown in \fig{fig:lgphase}. As we mentioned, the GFP splits the phase space into two parts. We note that the spacetime asymmetry traced by the coupling $W$ can take any values. Since our world is in the vicinity of the GFP, where the Lorentz symmetry is guaranteed, it seems appropriate to choose $W=1$ there. The exponents of the couplings are given by their mass dimensions, i.e. $s^{G}_{\lambda}=2$, $s^{G}_{g'}=-2$, $s^{G}_{c}=0$, $s^{G}_{W}=0$, 
Although the choice of $g'$ can nicely hide the effect of $W$, this coupling can give information about the Lorentz symmetry. The rescaling from $g$ to $g'$ can be considered as a rescaling of the Newton coupling in such a way, that the Lorentz symmetry is preserved during the evolution. If we turn back to the original coupling $g$ we expect the space phase structure to change, and it happens indeed. In \fig{fig:lgwp} we present a phase diagram in the 3-dimensional phase space, that includes $W$ as the vertical axis.
It can be clearly seen, that although $W$ seems to keep its value 1 in the beginning of the evolution, at a certain value of the RG scale it starts to decrease, it will be analyzed later in \fig{fig:wsym}. As a result, in the high UV limit $W\to 0$, furthermore it brings $g\to 0$, too. It implies that the standard form of the Reuter fixed point with finite values of the couplings at the fixed point cannot be maintained. One can interpret the horizontal sections of the figure as the position of the Reuter fixed point for each value of $W$. Eventually the Reuter fixed point is deformed due to the consequence of $W$ evolution. The Reuter fixed point in its traditional form freezes the evolution in the UV regime preventing the appearance of new physics around the Planck scale, furthermore the violation of the Lorentz symmetry at high energies. Our method and results can account for the latter issue by including the evolution of $W$. The migration of the Reuter fixed point due to $W$ makes us the possibility to find new relevant couplings, that can signal new physics there, since we can consider \fig{fig:lgwp} in such a way, that only the shadow of the Reuter fixed point exists due to the Lorentz violation, and the evolution towards the high UV region and it enables us to find new fixed points.

These facts show that the evolution of $W$ is crucial to understand the behavior of the model. The UV and the IR flows are presented in \fig{fig:wsym} for $W$ in the symmetric phase.
 We can see in \fig{fig:wsym} that independently of its initial values, the flow of $W$ decreases in the UV direction, and takes the vanishing UV fixed point value. This implies that the coupling dies out at high energies, implying that the time derivative term in the action also dies out. Therefore the Lorentz symmetry is violated in the UV region in such a manner that a static system emerges. According to the evolution of $W$ in \fig{fig:wsym} we can see, that the scaling of $W$ breaks down at around $t_l~0.5$, which can correspond to the scale of the Lorentz symmetry violation. The other couplings are almost constant at $t_l$, but not precisely, otherwise the evolution could be frozen there. This slight deviation from the constant fixed point value enables us to go beyond the Reuter fixed point scaling and reach the high UV regime. However, the memory of the traditional Reuter fixed point remains there (see e.g. \fig{fig:lgwp}) where, as a shadow, there is the traditional Reuter fixed point drawn by the opaque grey curves, that collect the trajectories, and then flow together to the high UV regime.
 
If we consider the evolution of $W$ in the IR limit in \fig{fig:wsym}, we can see that it keeps its initial value taken near the Gaussian fixed point, at least in the symmetric phase. It implies that the Lorentz symmetry is not broken in the IR, so the traditional RG treatment towards the IR limit keeps the Lorentz symmetry. This result shows that our Minkowski RG technique can preserve the Lorentz symmetry, although the regulator and the cutoff explicitly breaks it. The situation alters somewhat in the broken symmetry phase, although even then there occurs a long crossover region, where $W$ keeps its initial value at the Gaussian fixed point towards the IR. There occurs, however, another scaling region close to the singularity, that is discussed later.

\begin{figure}[H]
	\centering
	\includegraphics[width=0.45\linewidth]{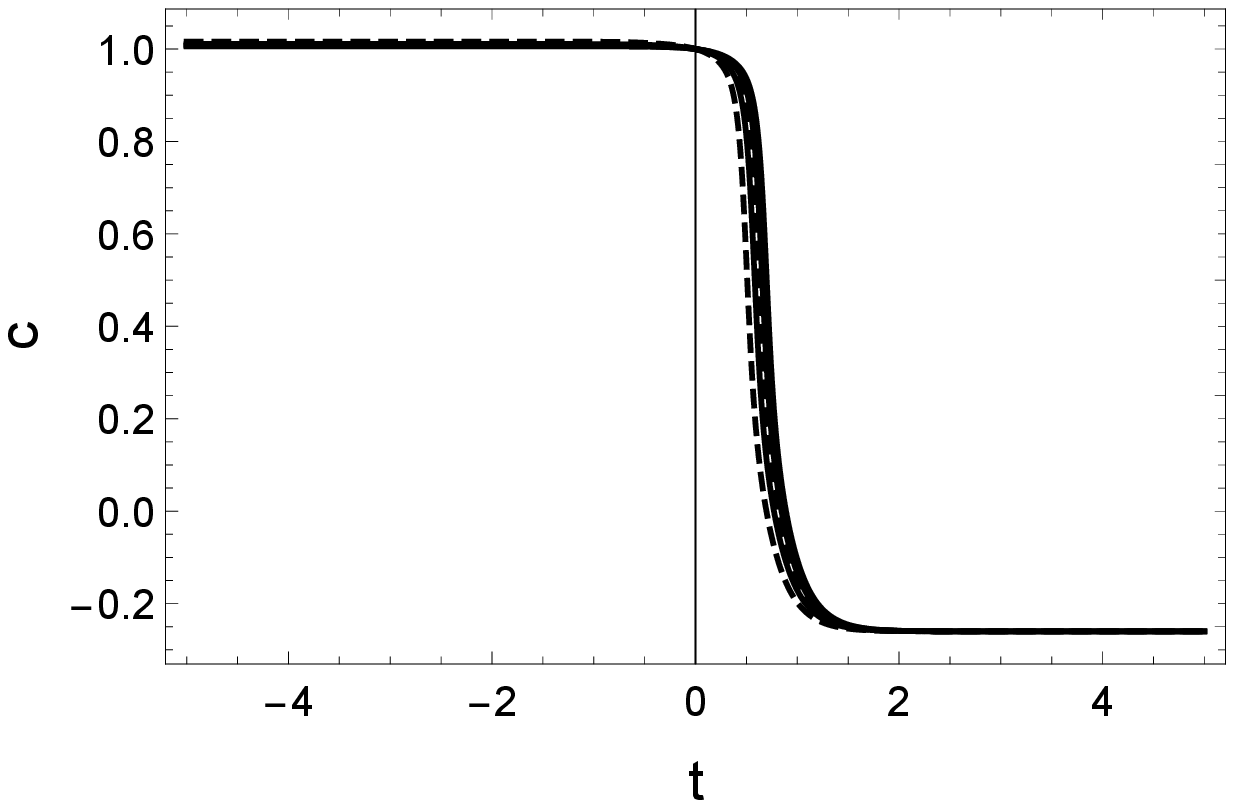} \includegraphics[width=0.45\linewidth]{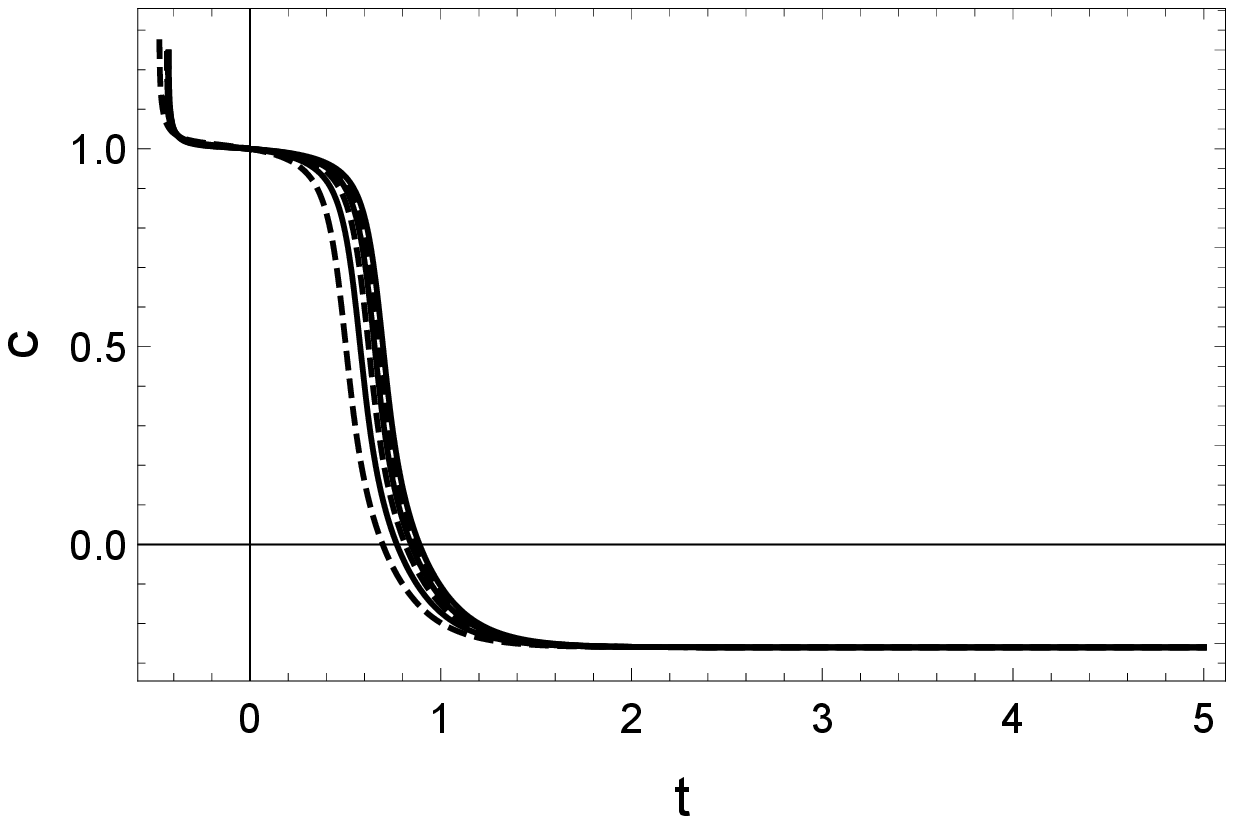}
	\caption{The flow of the coupling  $c$ is shown. The left figure presents evolution in the symmetric phase,
  with the initial value $c(k_i)=c_i=1$. The right figure shows the flow of $c$ in the broken symmetric phase. The different trajectories correspond to different initial values $W_i$, used in \fig{fig:wsym}.}
	\label{fig:csym}
\end{figure}

\begin{figure}[H]
	\centering
	\includegraphics[width=0.45\linewidth]{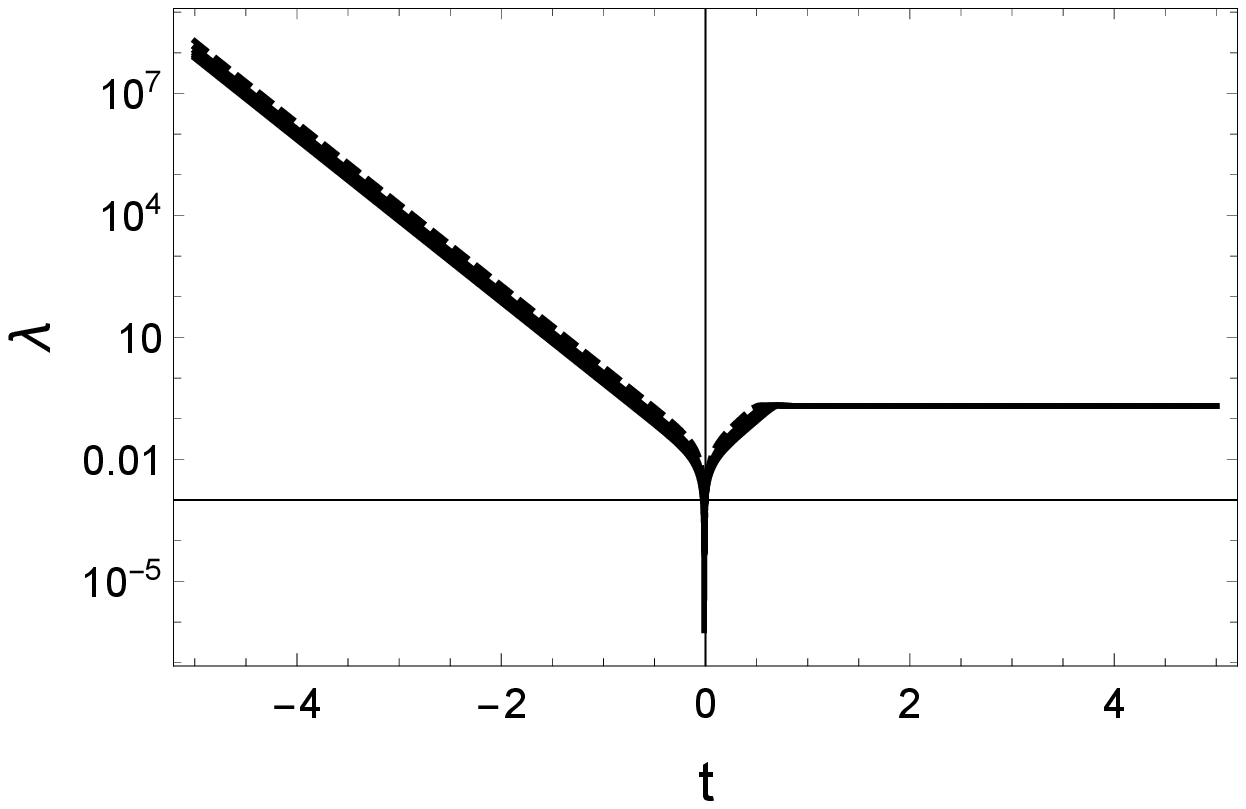} \includegraphics[width=0.45\linewidth]{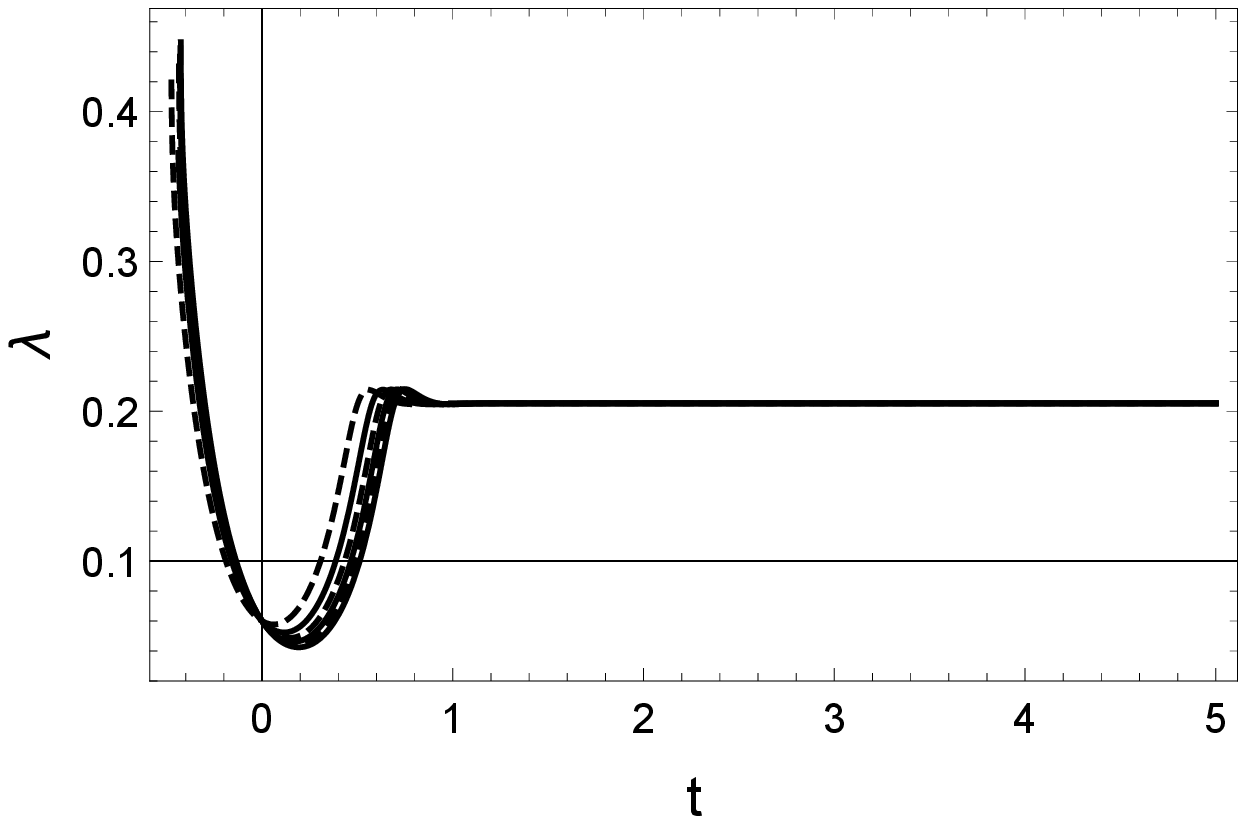}
	\includegraphics[width=0.45\linewidth]{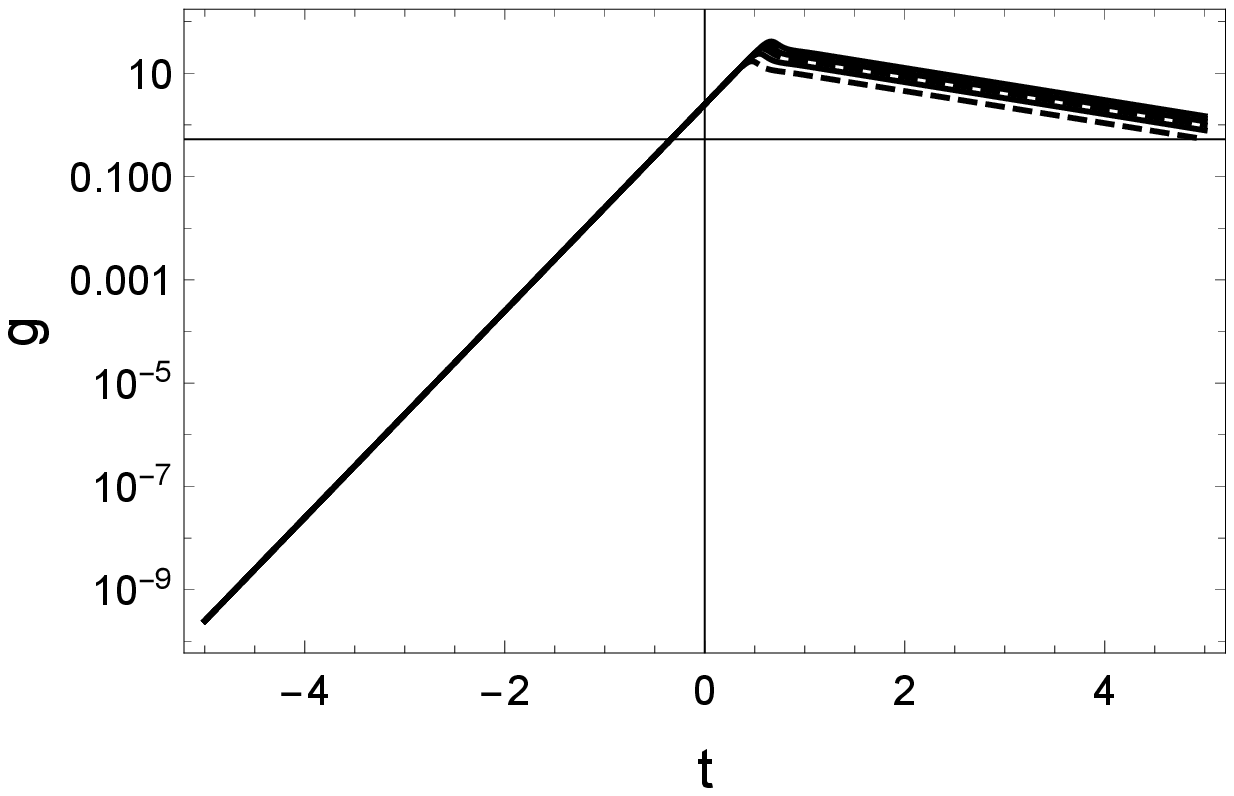} \includegraphics[width=0.45\linewidth]{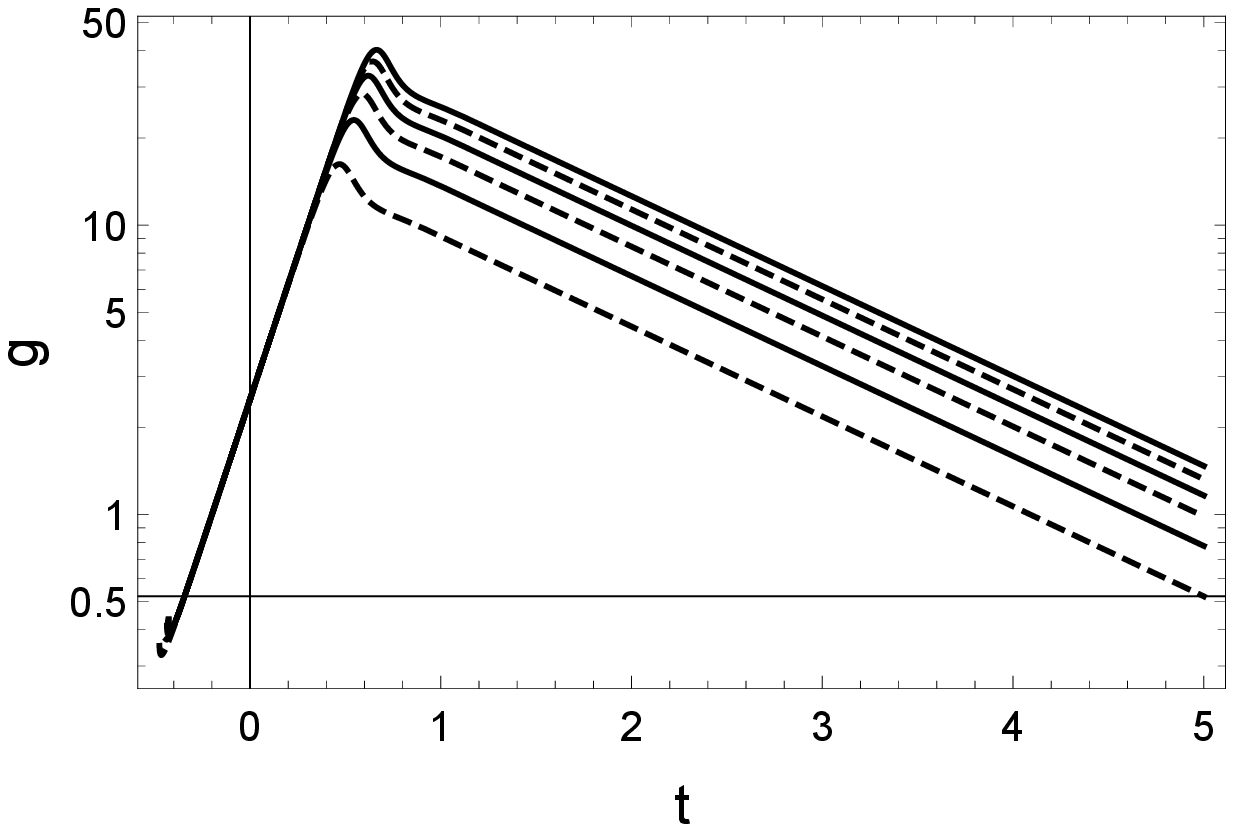}
	\caption{The flow of the coupling $\lambda$ and $g$ are shown in the symmetric phase (left figure)
and in  the broken symmetric phase (right figure). The different trajectories correspond to different initial values $W_i$, used in \fig{fig:wsym}.}
	\label{fig:lgsym}
\end{figure}

Next, we consider the coupling $c$  associated to the curvature term of the Einstein-Hilbert action. We demonstrated its UV and IR flows together in \fig{fig:csym}. In the UV region $c$ turns out to be relevant, and takes a constant negative value close to the Reuter fixed point, while it keeps its initial  value towards the IR. In the broken symmetric phase, another IR scaling law reveals itself once again. Although we were able to separate the term linear in the curvature $R$, and the associated coupling $c$ is relevant, but in the IR limit its effect is negligible.
For completeness, we show the scaling of $\lambda$ and $g$ in \fig{fig:lgsym}, where we see that both couplings go to finite positive values corresponding to the Reuter fixed point. Towards the IR region the trajectories belonging to the symmetric and broken symmetric phases behave  differently. Newton's coupling $g$ remains positive and decreases monotonically towards the IR, while the magnitude of the cosmological `constant' $\lambda$ goes to infinity in the IR region remaining positive in the broken symmetric phase and changing sign in the symmetric phase.
Therefore the flow of the couplings $\lambda$ and $g$ shows qualitatively the same well-known behaviour in our case, as it was obtained in the RG analysis of the 2-parameter Einstein-Hilbert theory. Thus we conclude that the flow of $\lambda$ and $g$ is not affected too much by the new couplings in the IR region.

\begin{figure}[H]
	\centering
	\includegraphics[width=0.45\linewidth]{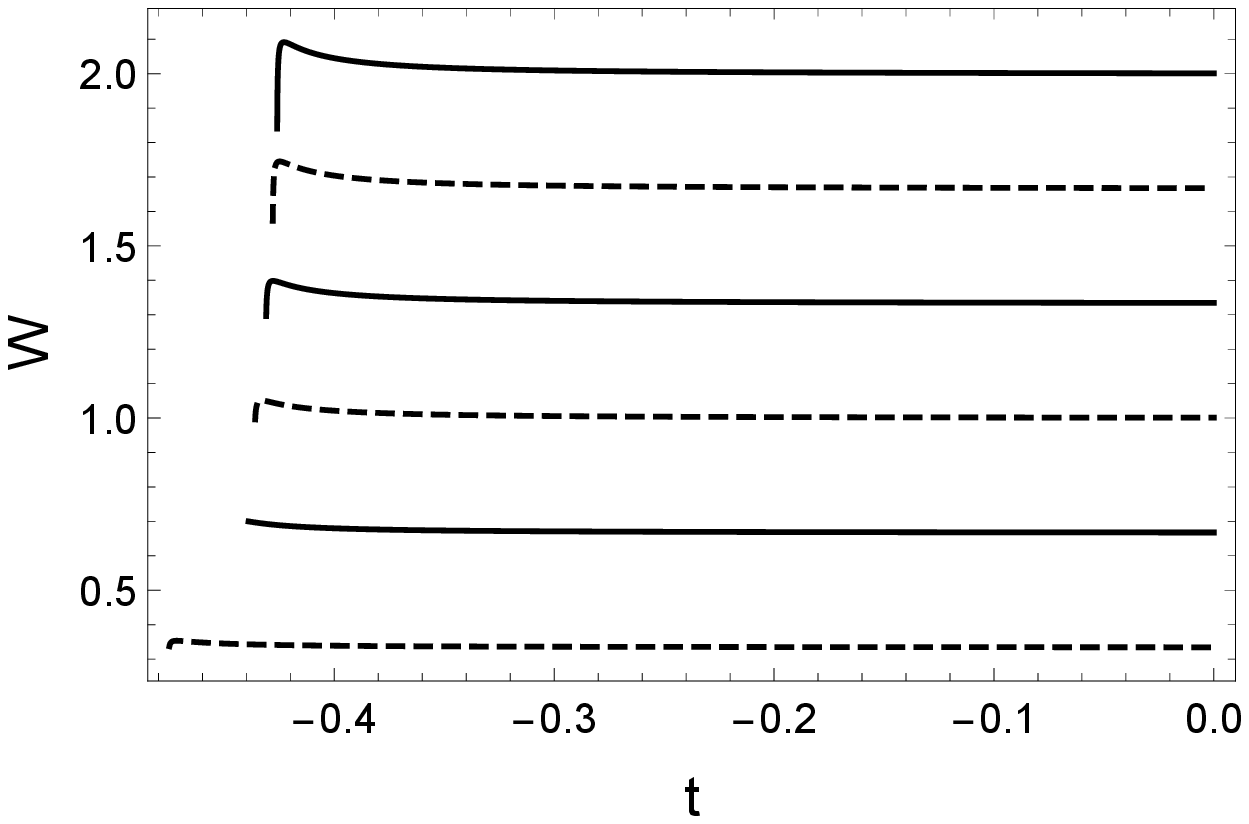} \includegraphics[width=0.45\linewidth]{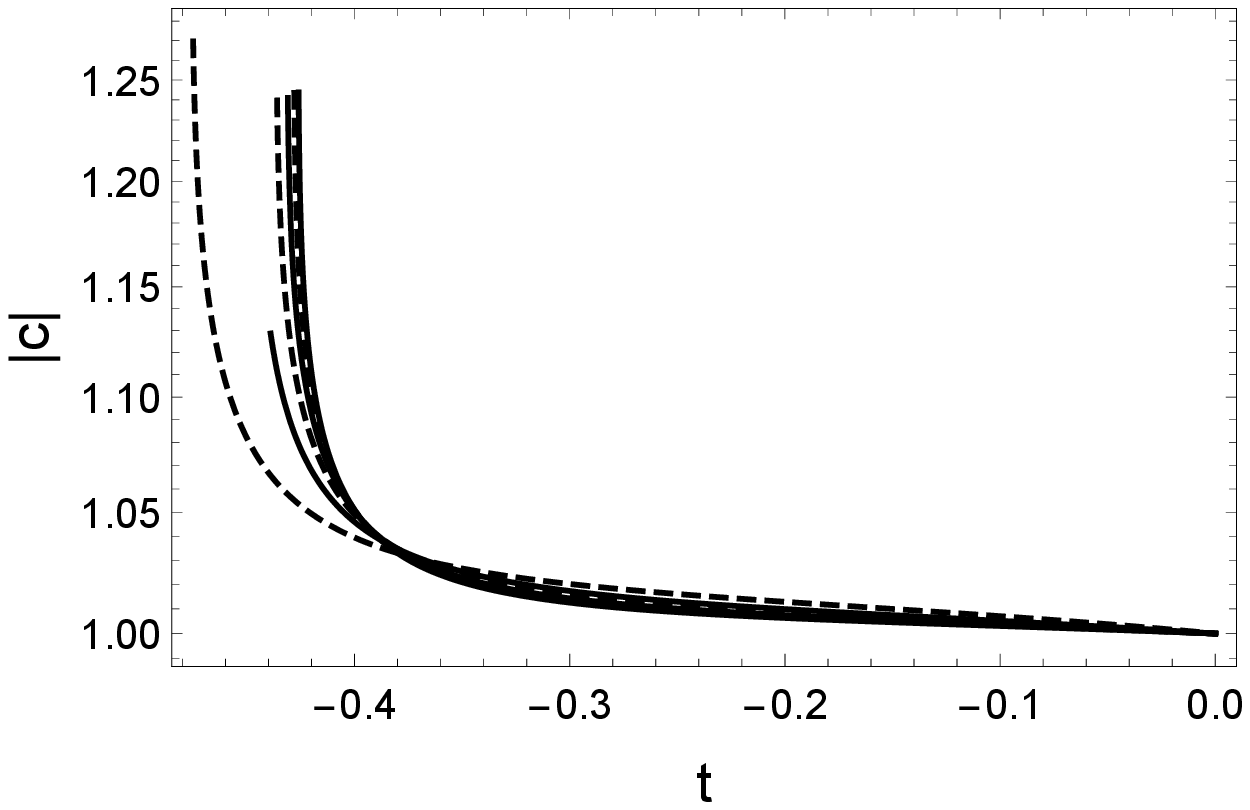}
	\includegraphics[width=0.45\linewidth]{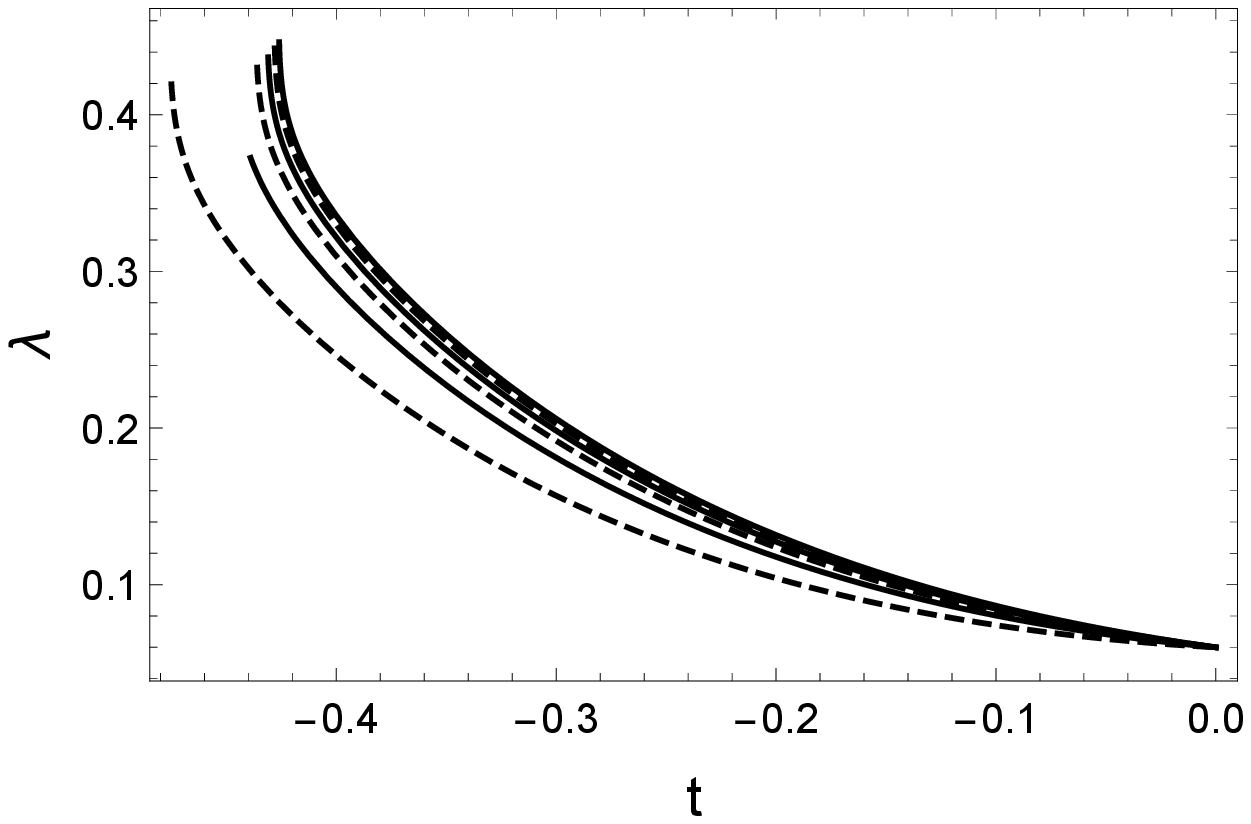} \includegraphics[width=0.45\linewidth]{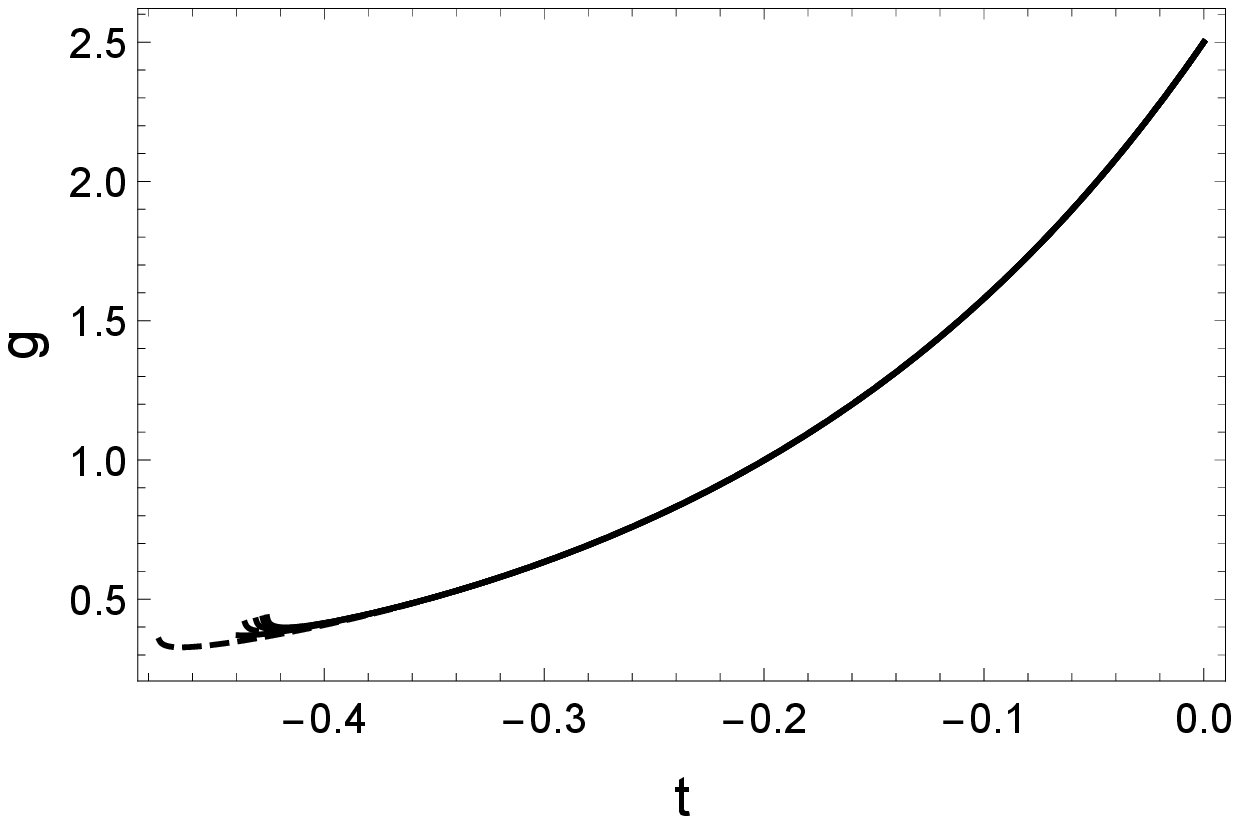}
	\caption{The RG flow of the various couplings is shown in the IR region in the symmetry broken phase. The initial values $W_i$ from bottom to top are: 1/3, 2/3, 1, 4/3, 5/3, 2, and $g_i= 2.5$, $\lambda_i=0.06$, $c_i=1$ for all trajectories.}
	\label{fig:brir}
\end{figure}

\begin{figure}[H]
	\centering
	\includegraphics[width=0.45\linewidth]{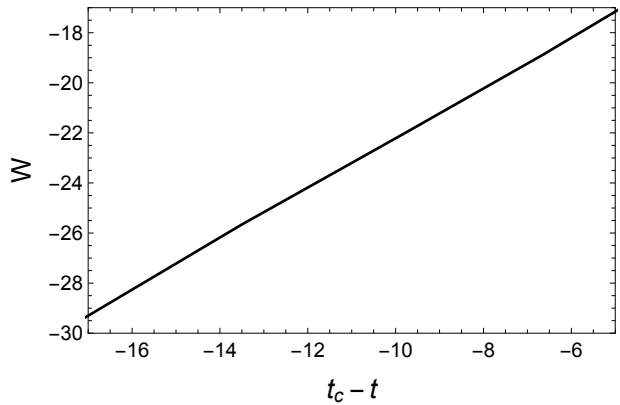}
	\caption{The scaling of the coupling $W$ is shown in the symmetry broken phase in the vicinity of the critical cutoff $t_c$.}
	\label{fig:wtct}
\end{figure}

Next we consider the short scaling region close to the singularity in the broken symmetric phase, shown in \fig{fig:brir}. First we note, that in the broken symmetric phase the RG trajectories cannot be followed up to arbitrarily small values of $t$, because there is a singularity in the beta functions, which appears at the pole of the propagator. Therefore, there is a finite critical cutoff scale $t_c$,  where the RG evolution stops. Its concrete value depends on the initial conditions, this is the reason, that various trajectories start to behave in a singular way at different values of $t_c$. All couplings start to diverge close to $t_c$, except $W$. By a proper rescaling of the couplings we can find, that close to the singularity the couplings have scaling behavior, suggesting that there is a fixed point in the IR regime of the broken phase. We made this analysis for the coupling $W$ and found that $W\sim 10^{\alpha_W(t_c-t)}$, where $\alpha_W=1$, see \fig{fig:wtct}. Similar type scaling behavior can be read off for the other couplings \cite{Braun:2010tt,Nagy:2012rn,Nagy:2012ef}. 

One can see from \fig{fig:wtct}, that the Lorentz symmetry is also violated in the deep IR region of the broken symmetric  phase. Since this phase seems to be physically more relevant (since the cosmological constant is positive in our world), it is interesting to notice, that around the GFP we have the Lorentz symmetry but in the high UV and in the deep IR regions it is violated. It seems that the breakdown of the applicability of the RG method around the singularity in the broken phase coincides with the violation of the Lorentz  symmetry. The singularity may suggest that new degrees of freedom are needed to describe the physical system. It is plausible to assume,  that the new modes do not possess Lorentz symmetry. We note, that low energy processes can be described by nonrelativistic tools, therefore the loss of the Lorentz symmetry there is not so surprising.

We can conclude that the RG blocking dynamically breaks the Lorentz symmetry in the UV limit in accordance with other results \cite{Eichhorn:2019ybe}. The symmetry breaking in the UV limit can have several reasons, but we should analyze our method used in this article in order to exclude it as a possible reason. We should admit that the choice of the Litim regulator breaks the Lorentz symmetry explicitly, and only the Callan-Symanzik scheme provides us Lorentz invariant regulator, however it cannot be used in quantum gravity, since it makes the 4-dimensional integrals divergent, and it requires further regularizations with cutoffs which can also break the Lorentz symmetry. We note, that in the UV limit, the Lorentz symmetry breaking of the Litim regulator becomes less and less strong, since the gliding cutoff scale goes to infinity, and the symmetry is restored in the $k\to\infty$ limit. This implies that the Lorentz symmetry violation is not sensitive to our choice of the regulator. Furthermore we emphasize that there is no problem in the IR direction, suggesting, that the symmetry breaking belongs to the Reuter fixed point, and does not occur due to the shortcomings of the RG technique.

\section{Summary}\label{sect:sum}

The Lorentz symmetry and its possible violation has been investigated in the conformally reduced asymptotically safe gravity by using the functional renormalization group method. Our calculations have been performed by a coordinate spacetime with Lorentz signature. This enabled us to introduce different field renormalizations for the time and the space directions, therefore the possible Lorentz symmetry violation can be revealed. Furthermore, we followed the evolution of the term linear in $R$ by a new coupling $c$.

We found that including these new couplings, the Reuter fixed point exists in the model. We showed, that the coupling $W$ that multiplies the time derivative is relevant, so as the coupling $c$. The exponents belonging to the Reuter fixed point are complex with negative real parts. We found that the coupling $W$ tends to zero in the UV limit, therefore the Lorentz symmetry is violated around the Reuter point. We also showed that in the high UV limit, where the Lorentz invariance violation becomes stronger and stronger, the Reuter fixed point migrates to the $g\to 0$ region of the phase space. In the IR direction the flow of the coupling $W$ behaves differently in the symmetric and the broken symmetric phases. In the symmetric phase the value of $W$ does not change in the IR region, thus no Lorentz symmetry violation occurs.

In the physically more interesting broken symmetric phase, however, there exists a critical scale where singularity occurs, and the constant crossover behavior of the coupling $W$ near the GFP is changed to another scaling law close to the critical scale, according to which $W$ tends to zero. This means that in the broken symmetric phase, there occurs Lorentz symmetry violation even in the deep IR regime. Although the coupling $c$ is relevant, it changes slightly during the evolution, and it decouples from the other couplings' evolution, therefore its effect is practically negligible.

It is yet an open question what causes the breaking of Lorentz symmetry near the Reuter fixed point. Can further terms, e.g. higher order or higher order derivative terms in the curvature restore the symmetry? The violation of Lorentz symmetry seems to be more natural in the deep IR region, nevertheless, it would be intriguing to investigate whether the functional  RG approach enables one the establish a kind of transition from relativistic to non-relativistic physics.

As a next step of this work, we intend to investigate the possible Lorentz symmetry violation in the original AS gravity model with its metric variable, but with the same technique, by handling the time and the space in a different way.

\bibliography{nagy}

\end{document}